\documentstyle{mn}

\input epsf
\def\plotone#1{\centering \leavevmode
\epsfxsize=\columnwidth \epsfbox{#1}}
\def\plottwo#1#2{\centering \leavevmode
\epsfxsize=.99\columnwidth \epsfbox{#1} \hfil
\epsfxsize=.99\columnwidth \epsfbox{#2}}
\def\plotthreev#1#2#3{{\centering \leavevmode
\epsfxsize=.99\columnwidth \epsfbox{#1} \hfil
\epsfxsize=.99\columnwidth \epsfbox{#2} \hfil}{\centering \leavevmode
\epsfxsize=.99\columnwidth \epsfbox{#3}}}

\def\apj{ApJ}                 
\def\apjl{ApJ}                
\def\apjs{ApJS}

\def\aap{A\&A}                
\def\aapr{A\&A~Rev.}

\def\mnras{MNRAS}

\def\nat{Nature}

\title[Soft state of Cygnus X-1]{Soft state of Cygnus X-1: stable disk and unstable corona}

\author[Churazov, Gilfanov \& Revnivtsev]{E.~Churazov,$^{1,2}$
M.~Gilfanov,$^{1,2}$ M.~Revnivtsev$^{2,1}$  \\
$^1$ MPI f\"{u}r Astrophysik, Karl-Schwarzschild-Strasse 1, 85740
Garching, Germany \\
$^2$ Space Research Institute (IKI), Profsouznaya 84/32, Moscow 117810, 
Russia}

\date{Accepted ????????????
      Received ????????????;
      in original form ????????????}

\pagerange{\pageref{firstpage}--\pageref{lastpage}}
\pubyear{2000}

\begin{document}
\maketitle

\begin{abstract}
Two component X--ray spectra (soft multicolor black body plus harder power
law) are frequently observed from accreting black holes. These components are
presumably associated with the different parts of the accretion flow
(optically thick and optically thin respectively) in the vicinity of
the compact source. Most of the
aperiodic variability of the X--ray flux on the short time scales is
associated with the harder component. We suggest that drastically different
amplitudes of variability of these two components are simply related
to the very different viscous time scales in the geometrically thin
 and geometrically thick parts of the accretion flow.   

In the geometrically thin disks variations of viscosity or mass
accretion rate occurring at large radius from the black hole on the
local dynamical or thermal time scales do not cause any significant
variations of the mass accretion rate at smaller radii due to a very
long diffusion time.  Any variations on the
time scales shorter than the diffusion time scale are effectively dampened. On
the contrary such variations can easily survive in the geometrically
thick flows and as a result the mass accretion rate
in the innermost region of the flow will reflect modulations of the
mass accretion rate added to the flow at any distance from the black
hole. Therefore if primary instabilities operate on the short
time scales then the stability of the soft
component (originating from the geometrically thin and optically thick
flow) and variability of the hard component (coming from the
geometrically thick and optically thin flow) are naturally explained.

For Cygnus X-1 overall shape of the power density spectra (PDS) in the soft and
hard spectral states can be qualitatively explained if the
geometrically thin disk is sandwiched by the geometrically thick corona
extending in a radial direction up to large distance from the 
compact object. In the hard state the thin disk is truncated at some distance
from the black hole followed by the geometrically thick flow. The
break in the PDS is then associated with the characteristic frequencies in
the accretion flow at the thin disk truncation radius.
\end{abstract}

\label{firstpage}

\section{Introduction}
A well known characteristic of accreting stellar mass
black holes is the presence of two drastically different spectral
states, first discovered in Cygnus X-1 by Tananbaum et al. 1972
(see Tanaka and Shibazaki 1996 for review). In the {\it Soft}
spectral state luminosity peaks at around 1 keV, while in the
{\it Hard} state luminosity is dominated by photons with the energy of the
order of 100 keV. This bimodality is believed to be related to very
different regimes of the accretion flow in the vicinity of the black
hole. Soft radiation is interpreted as a multicolor black body
emission originating from 
the optically thick (geometrically thin) disk (Shakura, Sunyaev 1973),
while hard emission, having a nearly power law shape at low energies
and a cutoff above $\sim 100$ keV, should come from an optically thin
and hot medium where Comptonization of soft seed photons by the hot
electrons plays an important role (e.g. Sunyaev and Tr\"{u}mper
1979). One of the popular models 
assumes that an optically thick disk is truncated at some distance from
the black hole and followed by an optically thin and hot flow 
(see e.g. Thorne and Price 1975, Liang and Price 1977, Esin,
McClintock and Narayan 1997, Meyer, Liu and Meyer-Hofmeister 2000). The hard
and soft spectral states of the source may then correspond to the situation
when an optically thick disk is truncated far from or close to the black hole
respectively. 

Another important property of the X--ray emission from black hole
candidates is a strong aperiodic variability on time scales longer
than 10 ms (see e.g. van der Klis 1994). The power density
spectra are very different during different spectral states suggesting
that the variability and spectral properties are closely linked. 

Cygnus X--1 is the best studied accreting black hole in our
Galaxy. Due to its brightness and persistent nature it was observed
virtually by every X--ray observatory flown to date. The source was
observed in different spectral states, also having very different
properties of the short time scale variability. We discuss below
qualitative model aimed to explain the changes in short time scale
variability correlated with changes in the spectral shape.

\section{Variability of the disk and corona}
\subsection{Constant and variable spectral components}
The state of the black hole candidates, when the spectrum contains a
strong soft component, are usually called ``High/Soft'' or ``Very High''
states (e.g. Tanaka and Shibazaki 1996). This soft spectral component has a
shape resembling multicolor black 
body emission and is believed to be produced by the standard optically
thick (geometrically thin) disk of Shakura and Sunyaev (1973)
type. Along with this ``black body'' emission a harder component is
often present in the spectrum, with an approximately power law
spectrum. Study of the variability properties of these two components  
led to the conclusion that most of the variability is associated with
the power law component and not with the ``black body'' emission (see
e.g. Miyamoto et al. 1994 for the GINGA observations of Nova Musca
1991). The same behavior have been observed for various sources 
manifested by the increase of the fractional RMS with energy. 
For the RXTE \cite{rxte} observations of Cygnus X--1 in the
soft state in June 1996 Gilfanov, Churazov and
Revnivtsev (2000) found that on the time scales shorter than 100
seconds the amplitude of the soft component variations is at least an
order of magnitude lower than that of the harder component. 

\begin{figure*}
\plottwo{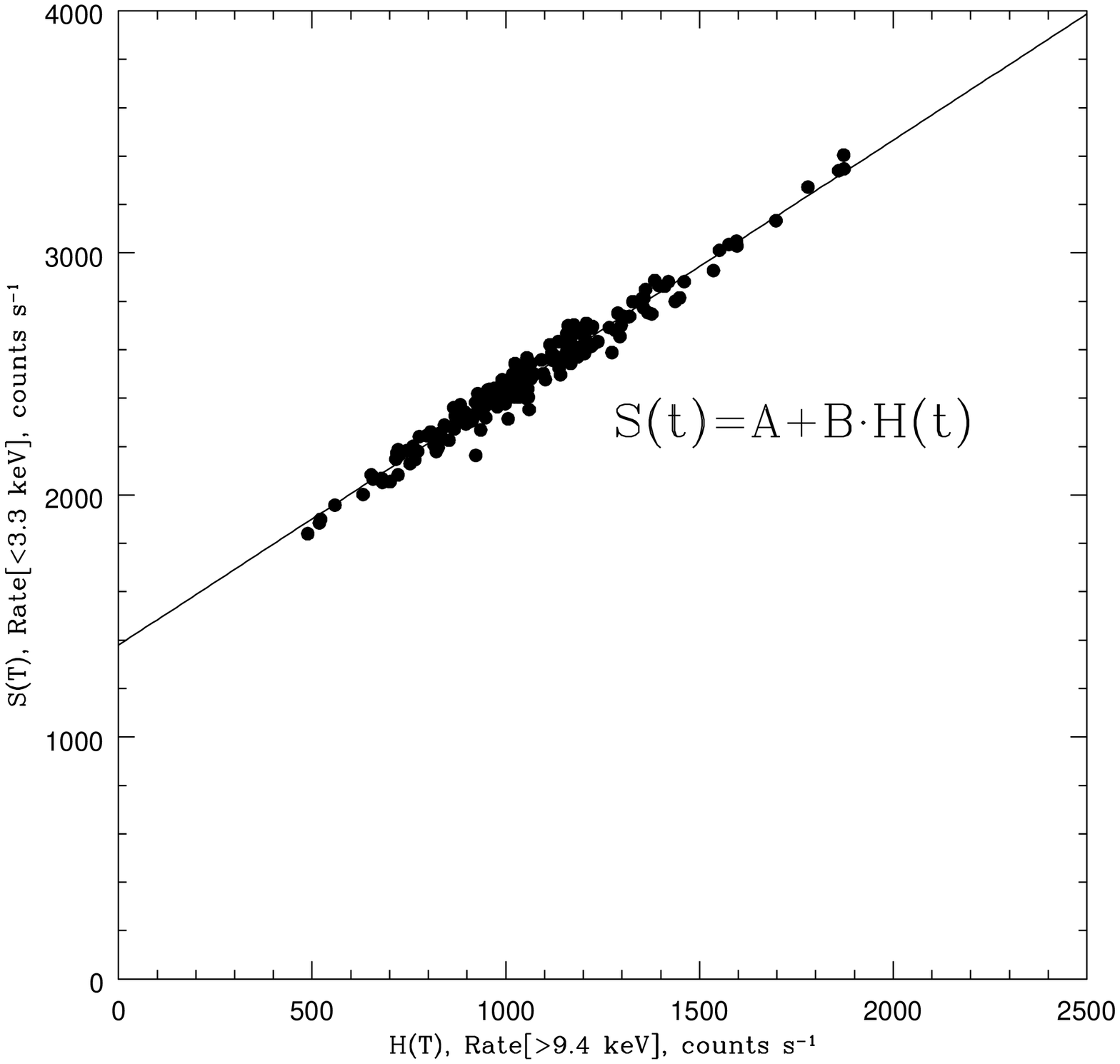}{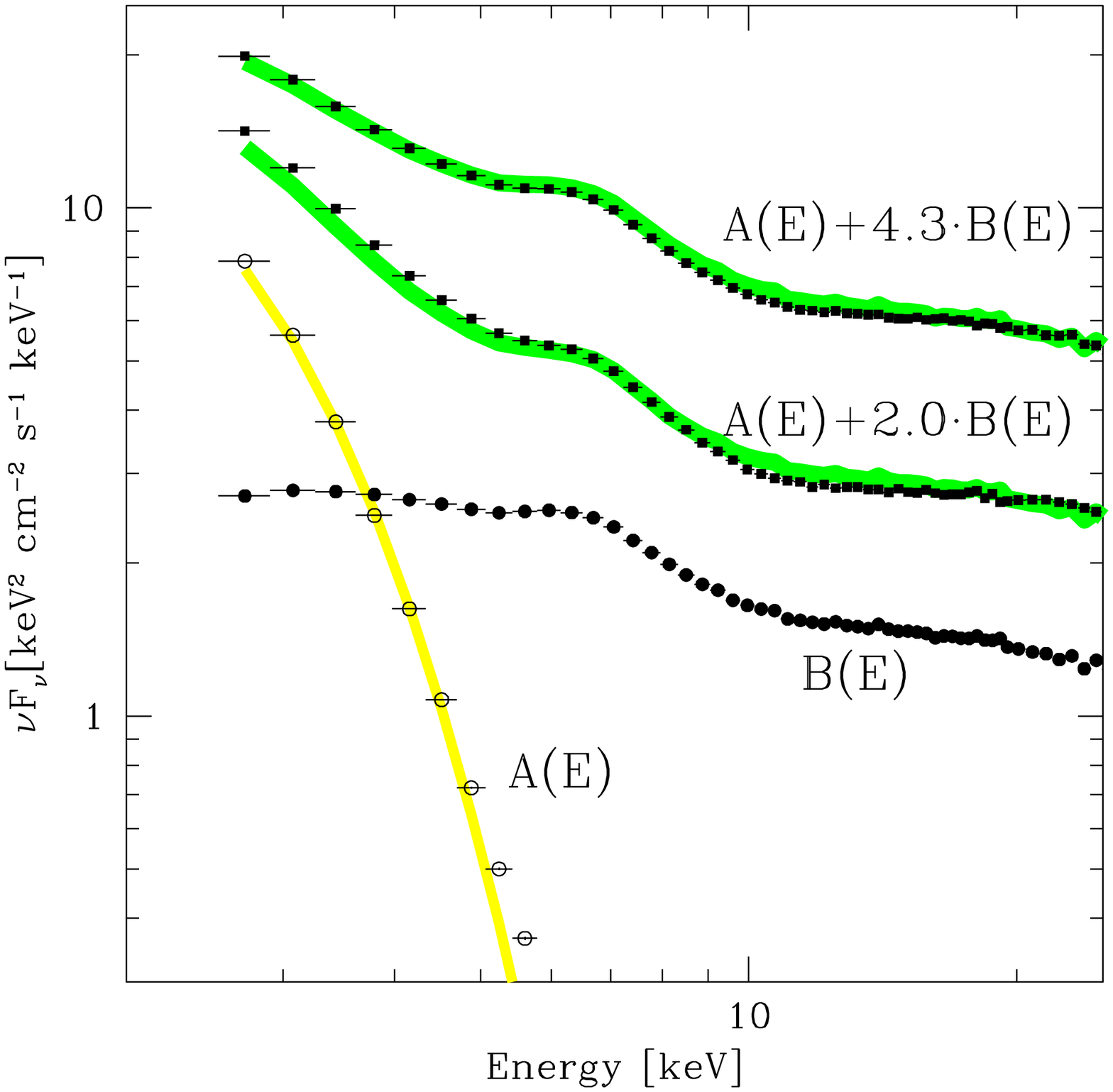}
\caption{{\bf a), Left:} The dependence of the count rate in the soft band
($<$ 3.3 keV) on the count rate in the hard band ($>$9.4 keV). Each
data point represent 16 s averaged count rate (RXTE observations on
1996 June 4 and 16). {\bf b), Right:} Spectra of ``constant'' (open
circles -- $A(E)$) and
``variable'' (solid circles -- $B(E)$) components, derived from the
linear fits of the correlation between count rate in different
channels. Normalization of the ``variable'' component $B(E)$ is
arbitrary. For comparison the 
light grey curve shows the spectrum of a multicolor black body
emission with a characteristic temperature of 0.5 keV. The two upper
spectra (solid squares) were averaged over the periods of time when
the  count rate above 9 keV was high and low respectively. The dark
grey lines show that these spectra can be reasonably well (within
10--15\%) approximated by a model $M(E)=A(E)+I*B(E)$ consisting from
the stable and variable spectral components  where $I$ (the
normalization of the variable component) is the only free parameter. 
\label{cor}
\label{spec}
}
\end{figure*}

For illustration we plot in Fig.\ref{cor}a the PCA/RXTE 
count rate in the soft band $S(t)=R(t,E<3.3~keV)$ as a function of
the count rate in the harder band $H(t)=R(t,E>9.4~keV)$. Here $R(t,E)$
is the observed (background subtracted) light curve in a given energy
band. The data points are the 16 s averaged values of the count rate
from the ``Standard Mode 2'' format of 
PCA. The observations were performed in 1996 on June 4 and 16. From
Fig.\ref{cor}a one can see that ({\bf i}) relative amplitude of variations
in the soft band  is a factor of $\sim 2$ smaller than in the hard 
band and ({\bf ii}) the correlation can be reasonably well
approximated by a linear relation $S(t)=A+B*H(t)$, where $A>0$. This
relation is shown in Fig.\ref{cor}a with a straight line. Similarly
good linear relation exists between the count rates in any pair of the
PCA/RXTE channels. Should the
observed spectrum vary in intensity only and not in shape, then one would
expect linear relation and $A=0$. The fact that the relation between the
count rates in two energy bands is close to linear, but $A$ is not
zero, means that additional stable component contributes to one or
both energy bands. For any two energy 
bands above $\sim 5$ keV the count rate in one band is
approximately proportional to the count rate in another band,
i.e. $A$ is close to zero.  
Very significant deviations of $A$ from zero level appear if one of
the selected energy bands is below 5 keV. This suggests that stable
component is present mostly at low energies. We then repeat the same
procedure of fitting the linear 
relation between $S(t)$ and $H(t)$ for every energy channel
setting $S(t)$ to the count rate in this channel and always
choosing $H(t)$ as a count rate above 9 keV. Thus
$S(t,E)=A(E)+B(E)*H(t)$, where $S(t,E)=R(t,E)$ and $H(t)=R(t,E>9)$ are the
observed light curves. The $H(t)$ therefore serves as a ``reference'' light
curve while $A(E)$ and $B(E)$ characterize respectively the
contribution of the  stable component to a given channel and the coefficient of
proportionality between the variations of the count rate in this channel
and the variations of the count rate above 9 keV. For the energy
channels above 9 keV there should be 
additional intrinsic correlation between $S(t,E)$ and $H(t)$
because $H(t)$ contains contribution from $S(t,E)$, but
experiments with various choices of the reference light curve $H(t)$
(e.g. using only narrow energy range instead of all counts above 9
keV) proved that resulting $A(E)$ and $B(E)$ vectors are relatively
insensitive to the choice of $H(t)$ as long as $H(t)$ contains only
contributions from energy channels above $\sim$ 5 keV.
The resulting vectors $A(E)$ and
$B(E)$ may be interpreted as the spectra (in units of ${\rm
counts~s^{-1} per~channel}$) of stable and variable
components. If the stable component is indeed present only at low
energies then with our choice of the $H(t)=R(t,E>9)$ the $A(E)$ will
reproduce both the shape and the normalization of the stable
component, while $B(E)$ will recover only the shape of the variable
component. The normalization of $B(E)$ depends on the particular
choice of the "reference" light curve $H(t)$.

These $A(E)$ and
$B(E)$ spectra have been approximately unfolded using the
XSPEC v10 \cite{ar96} command {\bf eeufspec} and assuming a power law
model with a photon 
index of $\sim$2.5. This procedure simply divides the observed
count rate in each channel by the effective area in this channel,
calculated for the assumed spectral model. The photon index of 2.5
approximately corresponds to the shape of the power law tail in the
source spectrum during June 1996 observations (Gierli{\'n}ski et al. 1999,
Gilfanov et al. 2000).  
The spectra of constant and variable components unfolded that way
are shown in Fig.\ref{spec}b.  The stable component $A(E)$ has a very soft
spectrum (open circles in Fig.\ref{spec}b). For comparison the light grey
line shows the spectrum of the multicolor black body disk emission
with the characteristic temperature $T\sim 0.5$ keV. The ``variable''
component $B(E)$ has a much harder spectrum (filled circles in
Fig.\ref{spec}b) and does not contain a strong soft component (see
Gilfanov et al. 2000 for details). Note that
although a good linear relation between count rates in any pair of
energy channels means that deconvolution in two components is accurate 
enough, these two spectral components may not necessarily have direct physical
meaning. However the fact that stable component closely reproduces the
multicolor black body disk emission both in terms of shape and
normalization (Gilfanov et al. 2000) strongly suggests
that stable component obtained from the above analysis simply
coincides with the multicolor black body disk emission.

Finally the two upper spectra shown in Fig.\ref{spec}b with solid
squares were obtained averaging 
the observed spectra over the periods of time when the  count rate
above 9 keV was high and low  respectively. These spectra contain
data points scattered over a period of time of more than 10 days when 
intensity of the source  in the hard band vary by at least a factor
of $\sim$4 (see Fig. 1a). The thick grey lines show
that these spectra can be reasonably well (within 10--15\%)
approximated by a model $M(E)=A(E)+I*B(E)$ consisting from the above
obtained stable and variable spectral components  where $I$ (the
normalization of the variable component) is
the only free parameter.  Thus from this analysis one can conclude
that all spectra observed by RXTE on 1996, June 4,16 during individual
16 seconds exposures can be reasonably well
approximated by a combination of a constant soft component and a
harder component which vary strongly in amplitude, but not much in shape.

\subsection{Power density spectrum in the soft state}
\begin{figure}
\plotone{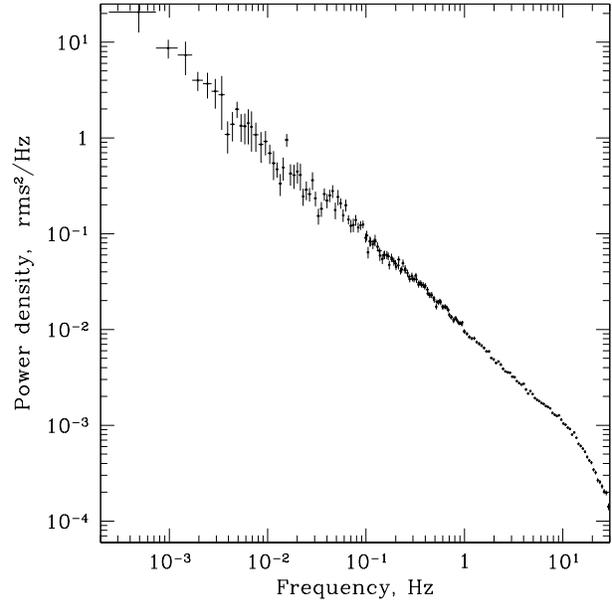}
\caption{ Power density spectrum of Cygnus X-1 during
June 1996 soft state (6-13 keV band, RXTE data). At lower energies the
shape of the PDS is the same, but the normalization is lower,
because of the dominant contribution of the stable soft component to
the observed X--ray flux.
\label{pdssoft}
}
\end{figure}

The typical power density spectrum (PDS) of the X--ray flux from Cygnus X-1
(in the 6--13 keV energy range) is shown in Fig.\ref{pdssoft}. The
PDS is calculated in units of squared fractional RMS per Hz, corrected
for the Poissonian noise contribution. In softer
energy bands the PDS has similar shape, but lower
normalization, because of the larger contribution of the stable soft
component to the source flux. The energy spectrum of Cygnus
X-1 in the soft state is discussed in details in Gierli{\'n}ski et
al. 1999 (see also Dotani et al., 1997 and Zhang et al., 1997). The
strong soft component was present in the Cygnus X-1 spectrum during these
observations (even in the RXTE band above $\sim$3
keV). The characteristic temperature of the black body component is
$\sim$0.5 keV (Gierli{\'n}ski et al. 1999) and this component provides
more luminosity than the power law tail. Assuming 
that the black body component is due to an optically thick disk and
that no strong advection is present in the optically thin flow one can
conclude that the inner radius of the disk 
is rather close to the black hole. Accurate determination of the 
disk inner radius  from the 1996 June 4,16 RXTE spectral data is
difficult because of the poor RXTE spectral response below 3 keV,
but the analysis of simultaneous RXTE and ASCA 1996 May 30 data (Dotani
et al. 1997, Gierli{\'n}ski et al. 1999) indeed indicates that the
inner radius  of the disk  is close to the black hole. We make a 
conservative assumption that the inner edge of the accretion disk
$R_{in}$ is within $\sim 20GM/c^2$, where most of the
gravitational energy of the accreting matter is released. 

The PDS of X--ray flux variations in this state
(Fig.\ref{pdssoft}) holds the same shape ($f^{-1}$ -- flicker noise)  from 
few $10^{-4}$ Hz up to 10 Hz, suggesting that the same physical mechanism
is responsible for the flux variations at all frequencies in this
range. Since these variations are associated with the power law
spectral component they presumably originate from an optically thin
region. Taking into account that $f^{-1}$ slope of the PDS holds
down to the frequencies of few $10^{-4}$ Hz it is difficult to imaging
that observed flux variations over this extremely broad
dynamic range of time scales (at least 4--5 orders of
magnitude) can  be provided by instabilities developing in the
innermost region within $R_{in}$. More promising seems the assumption
that the observed variability is due to instabilities occurring at
different distances (much larger than $R_{in}$) from the black hole
and then propagating into the innermost 
region, where the energy is released and X--ray photons are
produced. In particular Lyubarskii (1997) considered the fluctuations of mass
accretion rate in the innermost region associated with the
fluctuations of the viscosity parameter $\alpha$ in the accretion flow
at much larger radii. If the amplitude of $\alpha$
fluctuations at any radius is the same then variations
of the mass accretion rate through the boundary, placed at much
smaller radius, will have an $f^{-1}$ power density spectrum. Thus the
mass accretion rate is modulated at large distances from the black
hole, but the observed (modulated) X--ray flux is coming from the
innermost region. The broad dynamic range of the variability time
scales in the model of Lyubarskii (1997) can be provided by the broad
range of radii at which the viscosity is fluctuating.

\begin{figure*}
\plotthreev{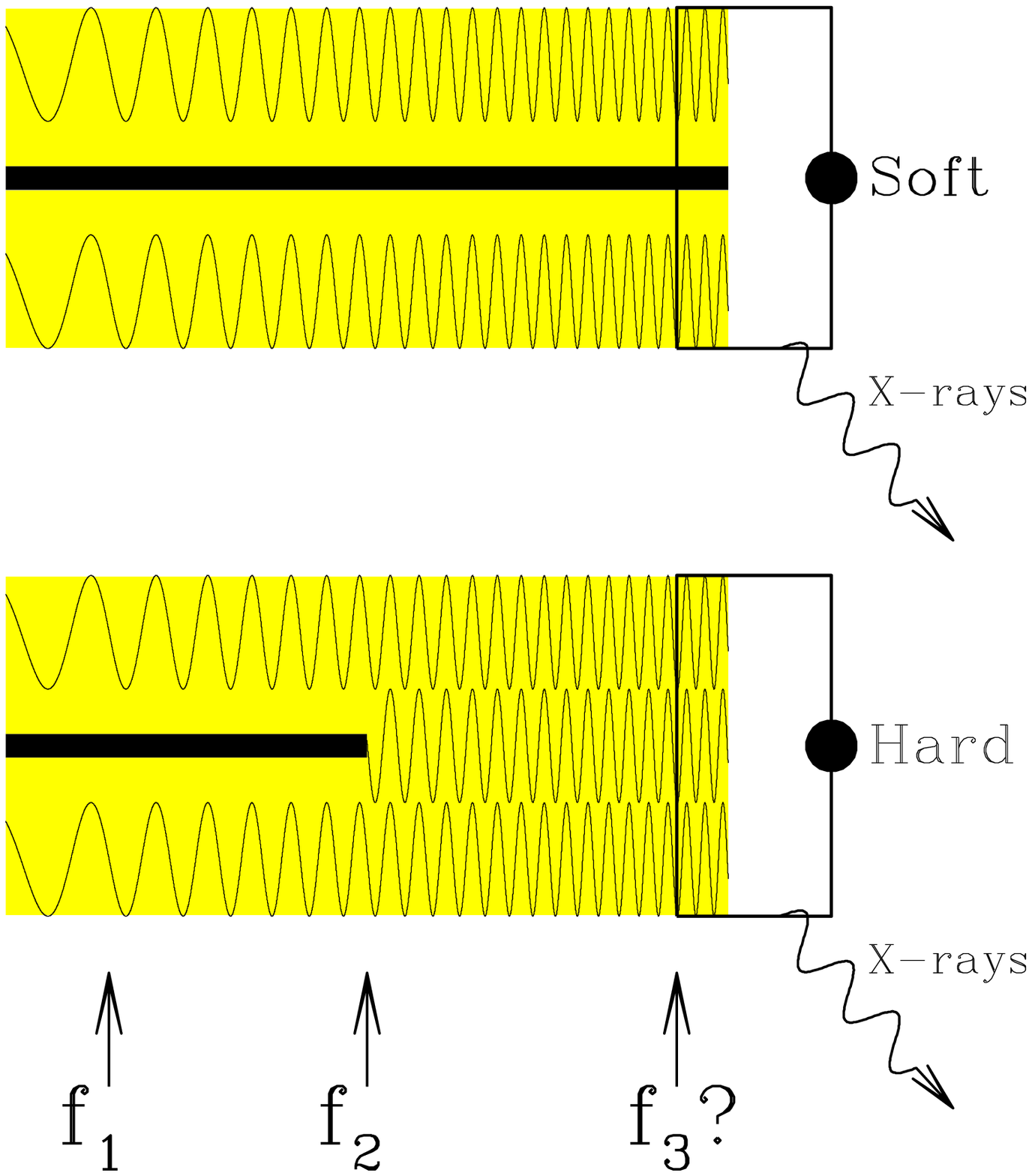}{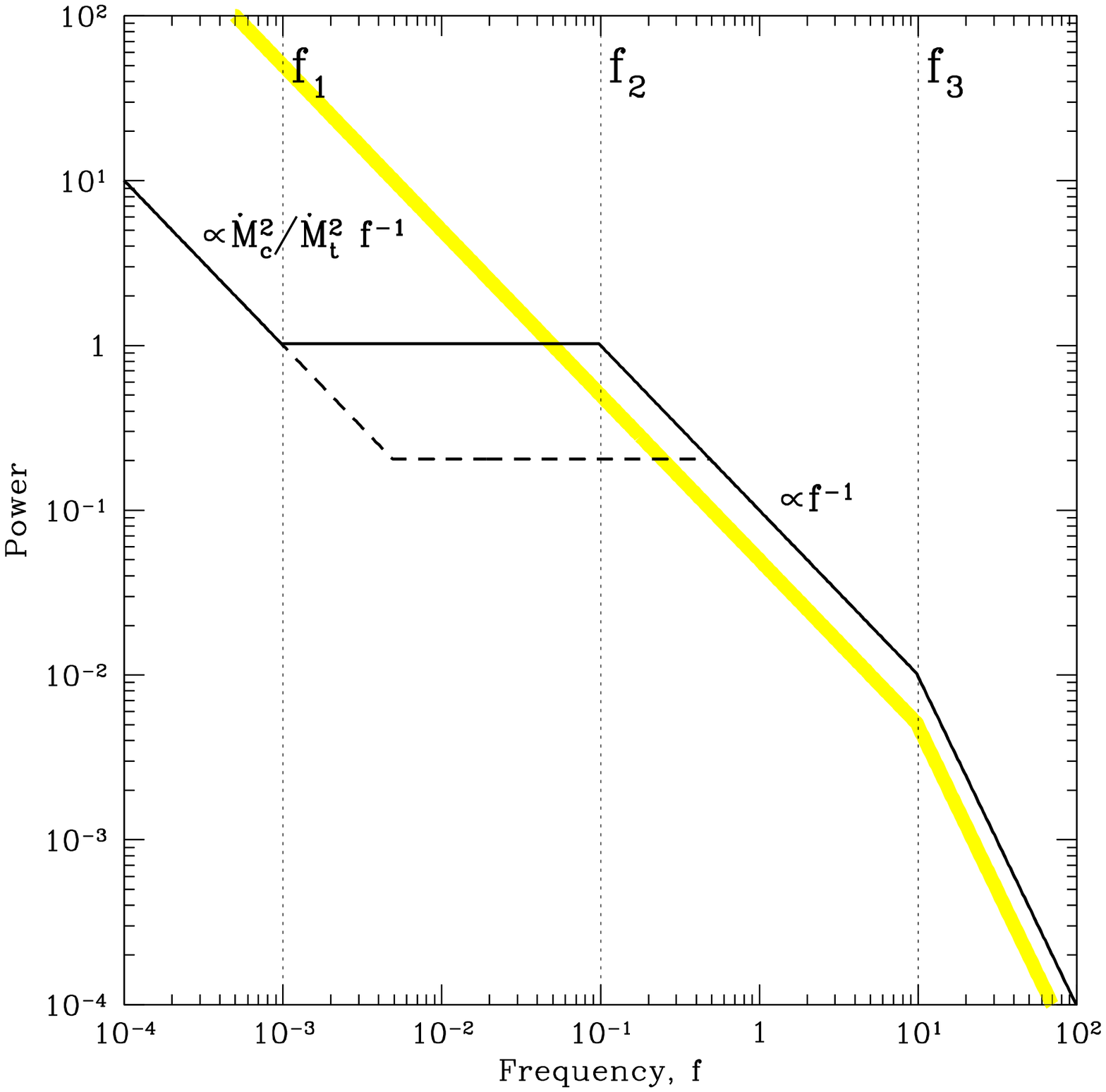}{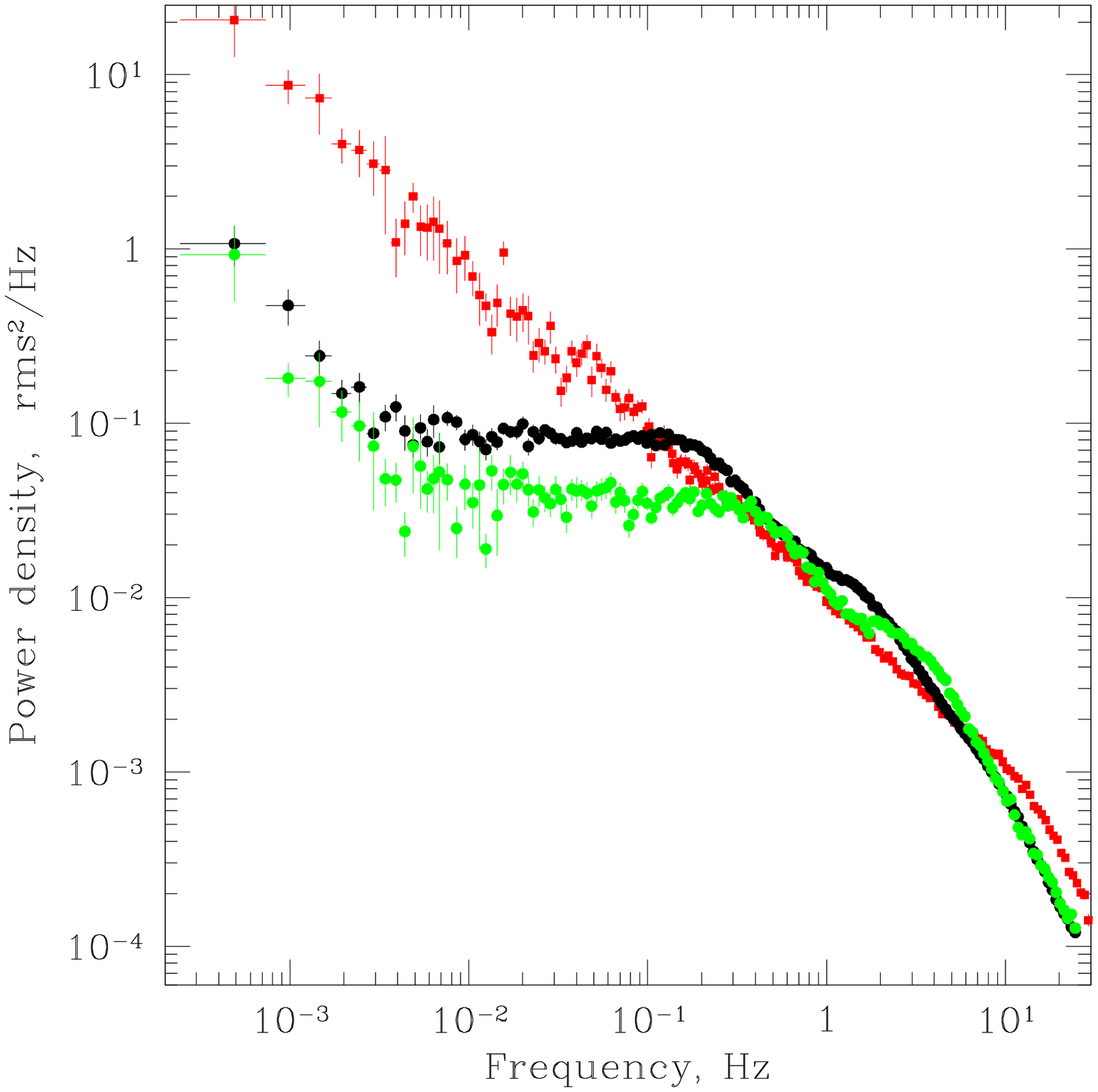}
\caption{{\bf a), Top--Left:} Sketch
of the adopted geometry for the soft and hard states 
of Cygnus X-1. The solid circle marks a position of a black hole. The
box shown by thin lines shows the area where most of the gravitational
energy is released and where most of the X--ray radiation is
emitted. The black ``slab'' shows the optically thick (geometrically
thin) accretion disk. In the soft state the inner edge of the disk is
close to the black hole, while in the hard state it is truncated far
from the energy release region. Sandwiching the disk
is an optically thin, geometrically thick corona (grey shaded regions),
extending in the radial direction up to a large distance from the
black hole. Oscillating curves show schematically that at different
radii the mass accretion rate in 
the corona is modulated on different time scales. This modulated
accretion flow reaches the innermost region and causes the
fluctuations of the observed X--ray flux over the broad range of the
time scales. 
{\bf b), Top--Right:} The overall shape of the PDS expected in the simple
geometry adopted here. In the hard state (thick solid line) there are
three breaks ($f_1$,$f_2$,$f_3$ shown by thin vertical lines) in the
power spectrum. $f_2$ is the characteristic frequency in the optically
thin flow at the disk truncation radius. Anticipated changes in the
power density spectrum associated with the inward motion of the disk
truncation radius are shown by the dashed line. In the soft state the power
spectrum (thick grey line) is a power law up to $f_3$. 
{\bf c), Bottom:} Typical power density spectra of Cygnus X-1 in
the hard (black and grey circles) and soft (squares) states. The
PDS are constructed from the RXTE data in the 6--13 keV energy range.
\label{sk1}
\label{sk2}
\label{pds}
}
\end{figure*}

Thus on one hand a stable optically thick disk extends down to a very
small radii $R_{in}$ (as indicated by the strong and stable soft
component) and on the other hand prominent variations  of
the harder component are present in a broad range of time scales (up
to at least $10^3$--$10^4$ s -- see Fig.\ref{pdssoft}), which are
likely to be associated with instabilities occurring at much larger
radii than $R_{in}$. The simplest explanation required to combine these 
two facts together is the assumption that along with a stable optically
thick accretion disk a variable (optically thin) corona
is present, which extends in radial direction up to large distances
from the black hole. The large extent of the corona in radial
direction is required in order to provide broad dynamic range of the
variability time scales. Various models involving
optically thin corona have been discussed in the
literature (e.g. Bisnovatyi--Kogan and Blinnikov 1976, Liang and Price 1977,
Galeev, Rosner and Vaiana 1979, Haardt and Maraschi 1991, Esin, McClintock and
Narayan 1997, Esin et al. 1998, Gierli{\'n}ski et al. 1999, see Poutanen
1998 for recent review).  The configuration of the 
accretion flow we adopted is schematically shown in Fig.\ref{sk1}a. 
Here the solid black slab shows the optically thick accretion disk,
which is assumed to be stable, sandwiched by an optically thin corona which is
shown as a grey shaded region. It is further assumed that
instabilities are operating in the corona which modulates the mass
accretion rate flowing through given radius. The "sine wave" with
varying period schematically shows that the time scales of modulations
in the coronal flow  increase with the radius. 
These modulations of the mass accretion rate are propagated down to
the inner region of the main energy release (shown by the thin box near 
the black hole) where the observed X--ray emission is produced. Thus
the variations of the observed X--ray flux coming from the inner
region of the main energy release reflect the modulations of the mass
accretion rate at much larger radii. In this
picture one would then expect to observe the spectrum consisting of
two components (soft stable component due to the disk emission and
harder variable component due to Comptonization in the corona). The
relative contribution of these two components to the luminosity of an
averaged spectrum would then reflect the ratio of the energy releases
in the disk and corona (or mass accretion rates if no strong advection
takes place in the corona). The PDS of the harder component may
then have the same shape over broad range of frequencies as indeed
observed (Fig.\ref{pdssoft}). The relative amplitude of the
X--ray flux variations should be lower at low energies where stable
multicolor black body component provides dominant contribution the
observed flux. 

\subsection{Power density spectrum in the hard state}
Let us now consider what kind of PDS we can expect in the hard state.
In the hard state the soft (black body) component is weak or absent. 
We assume below that that the optically thick disk is truncated at a
larger distance  from the black hole (so that emission from the disk
falls below X--ray regime). The adopted configuration of the accretion
flow is schematically shown  in Fig.\ref{sk1}a. Here the optically
thick (stable) disk ends at some radius and it is followed by an
optically thin flow which joins the coronal flow, which is present
above the disk. We assume that properties of this inner optically thin
flow in terms of amplitude and characteristic time scales of modulations of
the mass accretion rate are similar  to those of the corona. We
further assume that above the disk truncation radius larger fraction
of the mass accretion 
rate $\dot{M}_{d}$ is flowing through the thin disk rather then through
the corona $\dot{M}_{c}$. I.e. $\dot{M}_{c} \ll \dot{M}_{d}$. The total
mass accretion rate in the optically thin flow below the disk
truncation radius is $\dot{M}_t=\dot{M}_{d}+\dot{M}_{c}$. The PDS (the
black solid line  in Fig.\ref{sk2}b) is then expected to have several
distinct regions 
over frequency. These regions are schematically shown by the vertical
lines in the Fig.\ref{sk2}b labeled as $f_1$, $f_2$, $f_3$. The
corresponding geometrical regions in the accretion flow are marked
with the arrows in Fig.\ref{sk1}a.  

At high frequencies ($f \sim f_3$ in Fig.\ref{sk2}b) there is a
turnover of the PDS which may be due to the same reason as the
turnover in the soft state PDS (Fig.\ref{pdssoft}). The discussion of
this turnover is beyond the scope of this paper and we only briefly
speculate on it in Section 3. Detailed analysis of the high frequency
part of the Cygnus X--1 PDS is given in Revnivtsev et al., 2000. 

The range of frequencies from $f_2$ through $f_3$ is associated with the 
variability scales corresponding to the optically thin region of the
flow below the disk truncation radius (Fig.\ref{sk1}a). In this region
we might expect the flow to be similar to the corona flow in the soft
state and as a result the PDS should roughly reproduce the PDS in the
soft state (i.e. $Power \propto f^{-1}$ for $f_2<f<f_3$). The only
difference is that in the soft state additional soft component is
contribution to the observed flux. This soft component does not
contribute much at the energies higher than 5 keV and we therefore can
directly compare the shape and the normalization of the PDS in the
hard and soft state for the higher energy bands (Fig.\ref{pds}c). 
Here $f_2$ is a characteristic frequency of the modulations introduced by
instabilities operating in the optically thin flow at the truncation
radius of the disk. Below this frequency only a small fraction of
accreting matter contributes to variability  -- the fraction of matter
which goes through the corona. As a result normalization of the power
density spectrum, associated with the viscosity fluctuations in this
region should drop by a factor of $\sim \left
(\dot{M}_{c}/\dot{M}_{t}\right )^2$.  This component should
appear in the power density spectra as a $f^{-1}$ component at very
low frequencies as shown in Fig.\ref{sk2}b. The frequency below which
this component is dominating the PDS is marked as $f_1$. There is also
a transition region between frequencies $f_1$ and $f_2$. One can
expect the PDS to be flat in this frequency range. Indeed, the
variability of the X--ray flux in this range of time scales
is not associated with the ``locally'' induced fluctuations of the
accretion flow, which are suppressed by the factor $\sim \left
(\dot{M}_{c}/\dot{M}_{t} \right )^2\ll 1$, but rather with the
stochastic superposition of the modulations of the accretion rate
occurring at smaller radii. The frequency $f_1$, below which the PDS
switches from flat to $f^{-1}$ dependence can then be expresses as
$f_1=f_2\times\left (\dot{M}_{c}/\dot{M}_{t} \right )^2$. 

Thus the assumption of the variable corona sandwiching the stable disk
(as inspired by the soft state data -- Fig.\ref{pdssoft}) leads to a
prediction of a specific shape of the PDS in the hard state. For
comparison observed PDS in the hard state (for two RXTE observation in
March 1996) are shown in Fig.\ref{pds}c. Using the above arguments 
on the relative normalization of the two $f^{-1}$ regions of the PDS
one can estimate that in Cygnus X-1 the fraction of mass accretion
rate going through the corona above the disk is $\sim$
20\%. Variations in the PDS shape (in particular the shift of the
break frequency) is then interpreted as the 
change of the disk truncation radius and the related change of the
characteristic frequency. The expected change of the PDS caused by
inward shift of the disk radius is shown by the dashed line in
Fig.\ref{sk2}b. Similar behavior of the PDS (correlated change of the
break frequency and normalization of the band limited noise) was
first reported for Cyg X-1 by Belloni \& Hasinger (1990). 
When the disk truncation radius extends well down to
the innermost region the PDS switches from the ``3-breaks'' shape to
the ``1-break'' shape as observed in the soft state.

\section{Discussion}
Described above is a qualitative picture inspired by the variability
of the source during soft state. The suggested model is a
phenomenological one and we speculate below on the possible underlying
physics.

The first important question is why the mass accretion rate in the
thin disk would be stable, while in the corona it is variable. Note
that we do not address here the question of the accretion flow
stability (e.g. against thermal and viscous 
instabilities in the radiation-pressure-dominated part of the
disc)\footnote{Gierli{\'n}ski et al. 1999 argued that the disk was
stable during 1996 soft state of the source}. Instead one can think of
the effect on the mass accretion rate of the magneto-hydrodynamic turbulence,
which presumably serves as a source of the viscosity in the accretion flow
through the fluctuating magnetic stresses  (e.g. Hawley, Gammie and
Balbus 1995,  Brandenburg, Nordlund, Stein and Torkelsson 1995).

Lyubarskii (1997) considered the power density
spectrum arising from fluctuations of viscosity at different
radii which causes fluctuations of the mass accretion
rate. Fluctuations at one radius are related to the fluctuations at
smaller radii through the Green function of the diffusion equation
(Lynden--Bell \& Pringle 1974, Lyubarskii 1997). In his picture
fluctuations of viscosity at a given radius on the viscous time scales
$t_{visc}\sim \frac{1}{\Omega_K\alpha (H/R)^2}$ causes fluctuations of
the mass accretion rate at all smaller radii. Here $\Omega_K$ is a
Keplerian angular frequency at a given radius $R$, $\alpha$ is
viscosity parameter of Shakura \& Sunyaev 1973, $H$
is the half thickness of the disk. However because of the diffusive nature
of the disk accretion any fluctuations of the mass accretion rate on 
time scales much shorter than the 
diffusion time scale will not propagate towards
much lower radii (see Appendix), but instead will vanish in amplitude very
close to the radius at which they originated. E.g. if we
assume that actual fluctuations at a given radius are occurring at a
time scales $t_f$, comparable with the dynamical time scales
$t_d\sim  \frac{1}{\Omega_K}$ or thermal time scales $t_{th}\sim
\frac{1}{\Omega_K\alpha}$, then in the standard thin disk, where
$H/R \ll 1$,  we will always
have $t_f \ll t_{visc}$. Therefore such fluctuations 
will never propagate down to much smaller radii. Even if we
consider the fluctuations occurring in the inner zone of the
geometrically thin disk, which is emitting in X--rays (i.e. multicolor
black body component directly observed by X--ray telescopes) we can
easily show that the amplitude of
fluctuations on a time scale $\sim 
t_f$ will be significantly suppressed after propagating a distance $\Delta
R$ in a 
radial direction such as: $\frac{\Delta R}{R}\sim
\sqrt{\frac{t_f}{t_{visc}}}$. For larger $\frac{\Delta R}{R}$ the amplitude
of the mass accretion rate fluctuations will vanish.
E.g. for fluctuations on the thermal scales $\frac{\Delta 
R}{R}\sim \frac{H}{R}\sim 10^{-2}$ ($H/R\sim 10^{-2}$ is a typical
value for the standard geometrically thin disk, dominated by a gas
pressure). Therefore $N\sim\frac{R}{H}\sim 100$ 
different (``incoherent'') region over radius will contribute to the
observed flux, effectively suppressing the fluctuations by a factor
of $\sim \frac{1}{\sqrt{N}}\sim 10^{-1}$. Thus for the geometrically thin
disk fluctuations of $\alpha$ (or mass accretion rate) on the dynamical
of thermal time scales will not cause very prominent variations in the observed flux. 

On the other hand geometrically thick disks (e.g. ADAF flows) are much
more transparent for the high frequency oscillations. E.g. the fluctuation
of the viscosity ($\alpha$ parameter) at some radius (at the
dynamical or thermal time scales) will affect the mass accretion rate at
all smaller radii and thus may provide fluctuations of observed flux
coming from the innermost region of the accretion disk. Thus
qualitatively the apparent stability of the disk compared to the corona could
be understood as the result of a much longer diffusion time in the disk,
which suppresses propagation of fluctuations.

In the above model the characteristic frequencies are related to the
position of the disk truncation radius: the smaller the truncation
radius the higher the characteristic frequencies (in particular the break
frequency). At least qualitatively this trend is consistent with correlation
of the spectral and timing parameters observed in the black hole candidates.
E.g. for Cygnus X--1 and GX 339--4 the increase of the characteristic
frequencies correlates with the steepening of the spectra and the increase of
the reflected component (Gilfanov et al., 1999, Revnivtsev et al.,
2000),
which may be related to the increase of the cooling of the Comptonization
region by the soft photons from the disk and the increase of the
fraction of the 
hard flux intercepted by the thin disk as it approaches black hole.
In GRS1915+105 and XTE J1550--564 the QPO frequency, which in turn 
correlates with the break frequency, is well correlated with the soft component
flux (e.g. Trudolyubov et al. 1999, Muno et al. 1999).

Of course the above representation of the PDS as a 3-break function is
a gross oversimplification. In reality the shape of the PDS will be much more
complex. E.g. broad humps may appear near the break frequency because the
geometrically thin disk at the truncation radius supplies mass at a steady
rate to the ``unstable'' inner geometrically thick region. In such
conditions any increase of the accretion rate in the optically thin
region
must be followed by the decrease of the accretion rate at later
moments of times (since
total mass supply rate by the thin disk is constant). As a result a
broad QPO--like hump  may appear in the power density
spectrum (Vikhlinin, Churazov, Gilfanov, 1994).  

The truncation radius of the disk may fluctuate with time
and affect the observed X--ray flux. Therefore even if the thin disk 
does not propagate the fluctuations of the mass accretion rate (at the
frequencies higher than diffusion time) we may see fluctuations of the soft
flux due to fluctuations of the disk truncation radius. The only case when
we should expect the soft component to be very stable is when the disk
extends all the way down to the marginally stable orbit at $\sim 3 R_g$
(i.e. in the genuine soft state). In the ``transition'' state (i.e. when the
disk is close enough to the black hole) fluctuations of the soft
component due to disk truncation radius can be observed directly. In the
hard state (when the disk is presumably truncated at a large distance from the
compact source) emission of the thin disk is almost outside the X--ray band
and variations of the soft disk flux can be observed indirectly through the
influence of the soft flux on the Comptonization region. E.g. variations of
the soft flux entering the Comptonization region with a given optical
depth and energy release to the electrons will result in variations of both
the flux and slope of the Comptonized spectrum. 

The high frequency turnover of the PDS (at
about 10 Hz) may be related to the instabilities operating in the region of
the main energy release. Part of the gravitational energy has already  been
released (and emitted) at larger radii. Therefore the amplitude of
X--ray flux variations associated with the mass accretion rate fluctuations
added to the flow in the inner region may be suppressed.  Note that in the
case of  accretion onto a neutron star a large fraction of energy is
released at the neutron star surface. Therefore this turnover may be absent
in the power density spectra of the accreting neutron stars.

\section{Conclusions}
In the soft spectral state of Cygnus X-1, observed by RXTE in June 1996,
the black body component was remarkably stable while the harder (power
law like) component varied strongly on the frequency scales from 10 Hz
down to few $10^{-4}$ Hz. We suggest that such behavior
is due to presence of an optically thin corona above the optically
thick disk, which extends up to a large distance from the black
hole. The variations of the mass accretion rate (or viscosity)
occurring at a large distance from the compact object are propagated down
to the region of the main energy release. The reason for different
variability properties of the disk and corona (namely stable disk and
unstable corona) may be due to the fact that in the optically thick
(geometrically thin) disk any fluctuations at time scales shorter than
the 
diffusion time scales $t_{visc}\sim \frac{1}{\Omega_K\alpha (H/R)^2}$
are effectively dampened and are not propagated down to small
radii. The assumption that in the hard state the disk is truncated at some
distance from the black hole (larger than the last marginally stable orbit)
then naturally lead to an explanation of the overall shape of the
power density spectra of black hole candidates in this spectral state.

\section*{acknowledgments}
We are grateful to Philip Armitage, Yura Lyubaskii, Friedrich Meyer, Henk
Spruit and Rashid Sunyaev for useful 
discussions, Andrzej Zdziarski and anonymous referee for
helpful comments. This research has made use of data obtained through the High
Energy Astrophysics Science Archive Research Center Online Service, provided 
by the NASA/Goddard Space Flight Center. The work was done in the
context of the research network 
"Accretion onto black holes, compact objects and protostars"
(TMR Grant ERB-FMRX-CT98-0195 of the European Commission).
M.Revnivtsev acknowledges partial support by RBRF grant 97-02-16264
and INTAS grant 93--3364--ext.

\appendix
\section{Dampening the variations in the geometrically thin disk}
In this appendix we formally demonstrate the natural fact that in a
geometrically thin disk
variations of the viscosity at some radius on a time scale shorter than the
diffusion time scale causes negligible variations of the mass accretion rate
at smaller radii. 

Consider for simplicity a geometrically thin (gas pressure dominated) disk
with the constant ratio 
$\frac{H}{R}$, where $H$ is a half thickness of the disk and $R$ is the
distance from the compact object. Let us assume that viscosity ($\alpha$
parameter) suddenly increases in a narrow ring at some distance $R_1$ from
the center. Following Lyubarskii (1997) the deviation of the mass accretion rate
from the steady state value at the radius $R=r\cdot R_1$ is:
\begin{eqnarray}        \label{md1}
\dot{m}(r,t) \propto \frac{C^2}{t^2}\sqrt{r} e^{-\frac{C}{4t}(r^{3/2}+1)}
\times  \nonumber \\
\left [ I_{-2/3}\left (\frac{Cr^{3/4}}{2t}\right )-rI_{1/3}\left
(\frac{Cr^{3/4}}{2t}\right )  \right ]
\end{eqnarray} 
where time $t$ is expressed in units of $\frac{1}{\Omega_K}$ (here
$\Omega_K=\sqrt{\frac{GM}{R_1^3}}$ is a Keplerian frequency at the radius
$R_1$), 
$C\sim\frac{1}{\alpha (\frac{H}{R})^2}$, $I_\nu (x)$ is the Bessel function
of imaginary argument. The factor $C$ is of the order of $10^4$--$10^6$ for
the standard geometrically thin (optically thick) gas pressure dominated
disk and $C$ decreases when the thickness of the disk increases. 
In the limit of $r\ll 1$ (i.e. at the radii much
smaller than the radius $R_1$ where $\alpha$ is changing) the variations of
the mass accretion rate is obviously:
\begin{eqnarray}        \label{md2}
\dot{m}(t) \propto \frac{C^{4/3}}{t^{4/3}} e^{-\frac{C}{4t}}
\end{eqnarray} 

\begin{figure}
\plotone{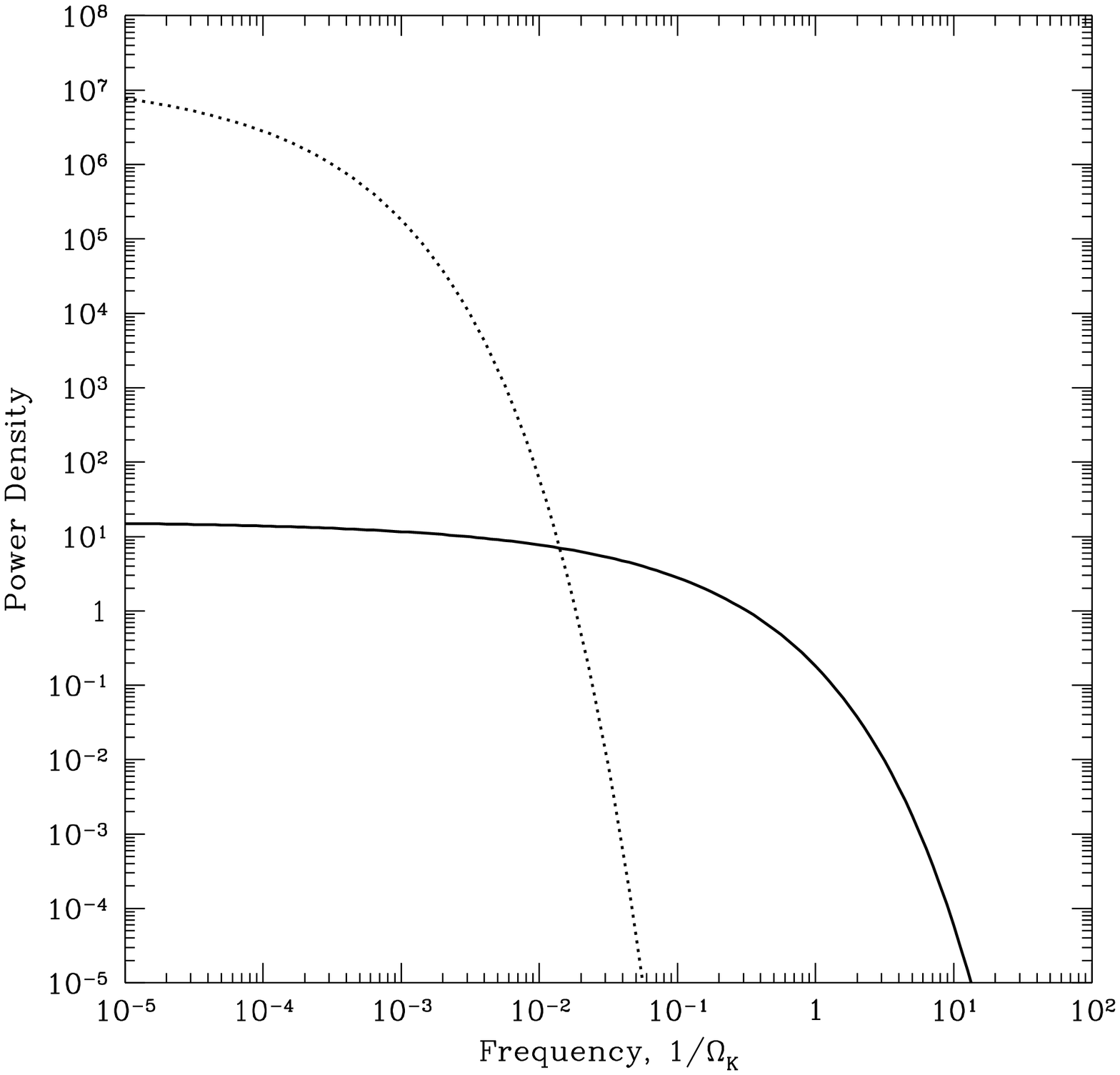}
\caption{The power density spectrum of the mass accretion rate variations at some radius $R$ caused by a sudden increase of viscosity ($\alpha$ parameter) in a
narrow ring at the distance $R_1\gg R$. Here
$\Omega_K=\sqrt{\frac{GM}{R_1^3}}$ is a Keplerian frequency at the radius
$R_1$. The dotted line correspond to the
case of a geometrically thin disk ($C\sim \frac{1}{\alpha (\frac{H}{R})^2}=10^4
$) and the solid line was obtained by formally setting $C$ to unity.}
\label{mpds}
\end{figure}

The power density spectra associated with the variability in the form
(\ref{md2}) are shown in Fig.\ref{mpds}. Here the dotted curve
corresponds to 
the case of $C=10^4$ (i.e. the typical value in the standard geometrically
thin disk). One can see that (as expected) virtually no variability is
present at high frequencies comparable to $\Omega_K$. For comparison the
solid line shows the power density spectrum formally calculated for
$C=1$\footnote{Note that $C\sim1$ necessarily means that disk is
geometrically thick and the above equations, derived in the limit of a
geometrically thin disk, are not applicable}. Here
significant variability is present up to high frequencies.

\label{lastpage}

\end{document}